\documentclass[letterpaper, 10 pt, conference]{ieeeconf}  
\IEEEoverridecommandlockouts                              
\title{\LARGE \bf
Impulsive Relative Motion Control   with Continuous-Time Constraint Satisfaction for  Cislunar Space Missions
}

\author{Fabio Spada$^{1}$, Purnanand Elango$^{2}$, Beh\c cet A\c c{\i}kme\c se$^{3}$
\thanks{\textsuperscript{*}This research has been partially sponsored by the Air Force
Office of Scientific Research (AFOSR) under grant FA9550-23-1-0646.}
\thanks{$^{1}$Ph.D. Student, William E. Boeing Department of Aeronautics
and Astronautics, University of Washington, Seattle, WA 98195.
        {\tt\small fspada46@uw.edu}}%
\thanks{$^{2}$Research Scientist, Mitsubishi Electric
Research Laboratories (MERL), Cambridge, MA 02139. P. Elango was at the University of Washington during the development of this work.
        {\tt\small elango@merl.com}}
\thanks{$^{3}$Professor, William E. Boeing Department of Aeronautics
and Astronautics, University of Washington, Seattle, WA 98195.
        {\tt\small behcet@uw.edu}}%
}

\usepackage[backend=biber,style=ieee,sorting=none, doi=false]{biblatex}
\usepackage{graphics} 
\usepackage{graphicx}
\usepackage[dvipsnames]{xcolor}
\usepackage{amsmath, amssymb} 
\usepackage{bm}
\usepackage{mathtools}
\usepackage{tikz}
\usepackage{threeparttable}
\usepackage{booktabs}
\usepackage{multirow}
\usepackage{svg}
\usepackage{subcaption}
\usepackage{array}
\usepackage{algorithm}
\usepackage{algpseudocode}
\usepackage{todonotes}

\newcommand{\submin}{\text{min}}
\newcommand{\submax}{\text{max}}
\newcommand{\nx}{{n_x}}

\newcommand{\na}{{n_a}}
\newcommand{\ti}{t_{\text{i}}}
\newcommand{\tf}{t_{\hspace{-0.02cm}f}}
\newcommand{\selector}[1]{^{#1}\hspace{-0.05cm}E}
\newcounter{problem}
\newcounter{remark}

\newenvironment{problem}
{ \par\vspace{0.2cm}\noindent 
    \refstepcounter{problem}
    \textbf{Problem \theproblem\, (P\theproblem)}%
  \par\vspace{-0.3cm}%
}
{}
{
    \color{magenta}
}
{}
\newenvironment{remark}
{\noindent 
\refstepcounter{remark}
\textit{Remark \theremark\,:}
}{}

\newtheorem{thm}{Theorem}
\newenvironment{pf}%
  {\par\noindent
   {\itshape\Elproofname}\enspace\ignorespaces}%
  {\par}
\def\Elproofname{Proof:}
\def\qed{\hfill$\Box$}

\newcommand{\ctscvx}{{\scalebox{1.1}{\textsc{{\scalebox{0.73}{ct-}}sc{\scalebox{0.73}{vx}}}}}}

\begin{document}

\maketitle
\thispagestyle{empty}
\pagestyle{empty}

\begin{abstract}

Recent investments in cislunar applications open new frontiers for space missions within highly nonlinear dynamical regimes. In this paper,  we propose  a method  based on Sequential Convex Programming (SCP) to loiter around a given target with impulsive actuation  while satisfying path constraints continuously over the finite time-horizon, i.e., independently of the number of nodes in which domain is discretized. Location, timing, magnitude, and direction of a fixed number of impulses are optimized in a model predictive framework, exploiting the exact nonlinear dynamics of non-stationary orbital regimes. The proposed approach is first validated on a relative orbiting problem with respect to a selenocentric near rectilinear halo orbit. The approach is then compared to a formulation with path constraints imposed only at nodes and with mesh refined to ensure complete satisfaction of path constraints over the continuous-time horizon. CPU time per iteration of 400 ms for the refined-mesh approach reduce to 5.5 ms for the proposed approach.

\end{abstract}

\section{INTRODUCTION}

This paper describes  and tests an optimal model predictive strategy to compute constrained relative trajectories between a chaser and a target spacecraft in nonlinear time-dependent dynamical environments, with  impulsive  controls. 
We propose to use a nonlinear model of the dynamics for the relative motion. 
The time-invariant Clohessy-Wiltshire model works well for eccentric orbits only when their eccentricity is low \cite{Yamanaka2002-ha} and is not accurate enough when strong non-Keplerian disturbances are present \cite{Yan2014-ct}. In multi-body scenarios, two-body models produce significant errors for propagation times similar to operational durations \cite{Franzini2019-al}. In all cases, the higher the distance between the chaser and the target, the less accurate the linearized model becomes. This ultimately motivates the use of the nonlinear relative dynamics model.

Chaser motion relative to target must satisfy safety constraints during operations. These include keep-out zones (KOZ) \cite{ISSP2019-hx} and approach corridors (AC) \cite{Goodman2006-cn}. Similarly, during inspection tasks, the chaser needs to stay within a keep-in zone (KIZ) close to the target \cite{Broschart2005-fe}. 

A simple KIZ constraint is satisfied when the chaser passively orbits its target on a  \textit{bounded relative orbit} (BRO). Initial conditions for BROs under zonal perturbations and for arbitrary eccentricities can be computed with both analytical \cite{Martinusi2014-cc} and numerical  methods \cite{He2019-dw, Dang2020-ea}. 
More generally, initial conditions of BROs form high-dimensional structures, known as \textit{tori}, that can be calculated in place of the initial conditions. This approach has for example been used to set up BROs around single non-spherical small-bodies \cite{Baresi2016-rm} and with two attractors in circular orbits around their center of mass \cite{Henry2021-ta}. 
\textit{Natural loitering} can also satisfy coarse KIZ requirements, but the trajectory that is passively flown depends on the considered dynamical environment \cite{DominguezUnknown-kv}: strict KIZ requirements need control. In \textit{forced loitering}, shooting is used to calculate the impulsive controls that enable periodic motion around the target \cite{Khoury_undated-vd, Sandel2024-ul}. However, impulsive controls and KIZ constraints are applied at pre-defined waypoints, thus no satisfaction of KIZ constraints is guaranteed along the coast arcs. \textit{Hovering} is similar to forced loitering: analytical methods are used for simplified dynamical models \cite{Broschart2007-fk}, whereas numerical schemes enable more elaborate constraints, as limitations on control authority \cite{Arantes-Gilz2019-fc, Sanchez2021-uq}. 
Vehicle motion can be constrained to surfaces \cite{Woodford2023-kj} if continuous control is available; semi-analytical methods have further been suggested to reduce computational load \cite{Zhao2024-jf}.
KOZ constraints are essential for safety; they are hence taken into account for both mission planning and control \cite{Petersen2023-gu}. 
    Graph-search techniques, such as Dijkstra's algorithm \cite{Frey2017-gw, Bucchioni2022-st}, and model checking algorithms \cite{Hibbard2023-dr} have been proposed to plan motion between disconnected sets. 

Past analyses address trajectory planning with polytopic KOZ as a mixed integer linear programming problem \cite{Richards2002-zj} and enable aggressive control by combining accurate KOZ models with robust nonlinear controllers \cite{Li2022-gr}.
Model predictive approaches can handle different maneuvers \cite{Weiss2015-kz}; KOZ models can be further tailored to prioritize one merit parameter over others \cite{Zagaris2018-pg}. Probability-based formulations \cite{Sanchez2020-mo} and backward reachable sets \cite{Aguilar-Marsillach2022-nk} enable satisfaction of safety constraints under modelled dispersions.

In this work, we solve the problem of bounded relative orbiting of spacecraft with impulsive control inputs by formulating it as an optimal control problem (OCP) with free timings of impulses. The solution approach that we propose extends the recently proposed {\ctscvx} framework \cite{Elango2024-dd} to handle state jumps due to impulsive control inputs. With generalized time dilation (GTD) \cite{Elango2024-dd,Kamath2023-id}, the OCP with free timing of impulses is cast into an equivalent OCP with fixed timings. Path constraints over the dense time horizon are reformulated as an equivalent set of isoperimetric constraints \cite{Elango2024-dd}. These two features improve performance of the optimization process: GTD allows to optimize both the location and timing of impulses; isoperimetric constraint reformulation ensures continuous-time constraint satisfaction without the need for mesh-refinement, which is necessary for safety-critical applications \cite{Richards2002-zj, Ranieri2008-ul} but time-consuming.
The problem is then discretized using multiple shooting, exactly penalized using $\ell_1$ penalty functions, and solved using \textit{prox-linear} method \cite{Drusvyatskiy2018-di}, a convergence-guaranteed sequential convex programming (SCP) algorithm. The contribution of this paper is threefold: firstly, for systems with jumps in the states, we prove equivalence between satisfaction of path constraints over a continuous-time horizon and satisfaction of a finite set of isoperimetric constraints; secondly, we test the outlined framework to solve a controlled bounded relative orbiting problem, considering a target on a southern $L_2$ near rectilinear halo orbit (NRHO) and both KOZ and KIZ constraints. The location and timing of impulses are optimized to maximize residence time in the permitted region. Both a two-impulse strategy with initial coasting and a three-impulse strategy are tested. In third instance, we compare our algorithm with a mesh-refinement-based strategy; this imposes path constraints only at nodes and ensures continuous-time constraint satisfaction by adding nodes. We demonstrate that our proposed algorithm reduces the computational time per iteration by nearly two orders of magnitude.

Following the notation in this section, the OCP is outlined in Sec. \ref{sec:def}. Time dilation, constraint augmentation and the employed SCP framework are detailed in Sec. \ref{subsec:dilation}, \ref{subsec:constr}, \ref{subsec:fin_SCP} respectively. The case study, corresponding numerical results and the comparison with the refined-mesh approach are respectively presented in Sec. \ref{subsec:proboutline}, \ref{subsec:results}, \ref{subsec:meshrefinement}. Conclusions are finally drawn in Sec. \ref{sec:concl}.
\subsection{Notation}
$\mathbb{R}$, $\mathbb{R}_+$, $\mathbb{R}_{++}$ denote the full set, the nonnegative subset and the positive subset of real numbers. Vectors are denoted with boldface notation; $\selector{(\cdot)}$ is such a matrix that $\selector{(\cdot)}\bm{x}=(\cdot)$. $n_{(\cdot)}$ is the dimension of $(\cdot)$. Given $\mathcal{D}\subset\mathbb{R}$ defined as $\mathcal{D}\coloneqq\bigcup_{i=1}^n\mathcal{D}_i$ and the continuous variable/function $(\cdot)$ defined on $\mathcal{D}$, we indicate with $(\cdot)_i$ the restriction of $(\cdot)$ to $\mathcal{D}_i$; furthermore we denote with $(\cdot)^{-}_i,(\cdot)^{+}_i$ the values of $(\cdot)_i$ at the left and right endpoints of $\mathcal{D}_i$. Component-wise inequalities are indicated with the symbols $\preceq, \succeq$.

\section{PROBLEM DEFINITION}
\label{sec:def}
\begin{figure}[h!]
\input{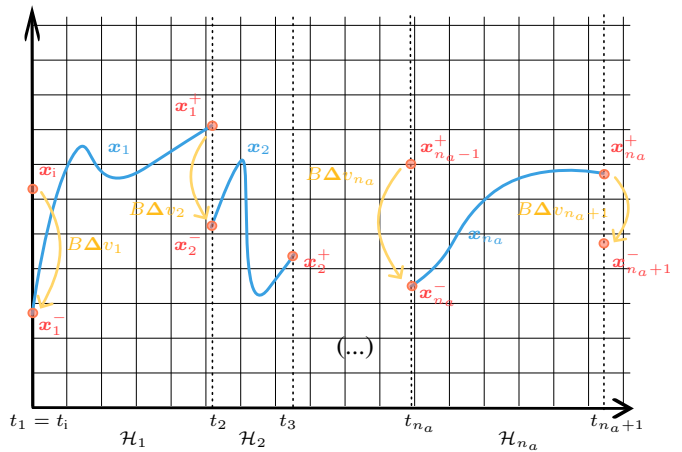}
\caption{One-dimensional representation of the considered dynamics with impulsive controls.}
\label{fig:repr}
\end{figure}

Let us consider the time horizon $\mathcal{H} \coloneqq [\ti, \tf]$, with fixed $\ti$ and variable $\tf$. Let us divide $\mathcal{H}$ into $\na$ arcs, indexed by the set $\mathcal{I}^{\na}$, with $\na+1$ time instants, indexed by the set $\mathcal{I}^{\na+1}$. The $i$\textsuperscript{th} arc $\mathcal{H}_i$ is defined as $\mathcal{H}_i \coloneqq [t_{i}, t_{i+1}]$, where endpoints are free; moreover, $\ti = t_1 < \dots < t_{\na} < t_{\na+1}$, where we replace $\tf$ with $t_{\na+1}$ to avoid duplicate variables. An impulsive control $\bm{\Delta v}_i \in \mathbb{R}^{n_v}$ is applied at each $t_{i}$. The relative-to-target state of the chaser, the absolute state of the chaser and the absolute state of the target are respectively described by the state vector $\bm{x}_i \in \mathbb{R}^\nx$, the vector $\bm{X}_i \in \mathbb{R}^\nx, i \in \mathcal{I}^{\na}$ and the assigned function $\bm{X}_{\text{T}}: \mathcal{H} \rightarrow \mathbb{R}^{n_x}$; therefore $\bm{x}_i \coloneqq \bm{X}_i - \bm{X}_{\text{T}}(t)$. \\The continuously differentiable function $\bm{f}^a: \mathbb{R}^{n_x}\times \mathcal{H} \rightarrow \mathbb{R}^{n_x}$ models the time-dependent absolute dynamics right-hand side (RHS). Given target initial state $\bm{X}_{\text{T,i}} \in \mathbb{R}^\nx$, the target state satisfies 
\begin{equation}
    \dot{\bm{X}}_{\text{T}}(t) = \bm{f}^a\left(\bm{X}_{\text{T}}(t), t\right), \; t \in \mathcal{H}, \quad \bm{X}_{\text{T}}(\ti) = \bm{X}_{\text{T,i}}
\end{equation}
Henceforth, we omit the dependency of states on $t$.
The relative dynamics of the chaser with respect to target, indicated with $\bm{f}: \mathbb{R}^{n_x}\times \mathcal{H}_i \rightarrow \mathbb{R}^{n_x}$, reads
\begin{equation} 
\bm{f}(\bm{x}_i, t) = \bm{f}^a(\bm{X}_{\text{T}} + \bm{x}_i, t) - \bm{f}^a(\bm{X}_{\text{T}}, t)\quad \left|\begin{array}{l} t \in \mathcal{H}_i, \\ \forall \, i \in \mathcal{I}^{\na} \end{array}\right.\end{equation}
Dynamics are uncontrolled  along each arc. \\Chaser relative initial state is fixed to $\bm{x}_{\text{i}} \in \mathbb{R}^\nx$; introducing an additional variable vector, conveniently denoted as $\bm{x}^-_{\na+1}\in \mathbb{R}^{\nx}$, the boundary conditions of the problem read
\begin{equation}
\arraycolsep=1.4pt
\begin{array}{rcl}
\label{eq:BCs}
\bm{x}^-_{1} &=& \bm{x}_{\text{i}} + B \bm{\Delta v}_{1}\\[0.5ex] 
\bm{x}^-_{i+1} &=& \bm{x}^+_{i} + B \bm{\Delta v}_{i+1}\quad \quad i \in \mathcal{I}^{\na}\end{array}
\end{equation}
where the \textit{allocation matrix} $B \in \mathbb{R}^{\nx \times n_v}$ allocates the components of $\bm{\Delta v}_i$ in the corresponding components of the state vector. 
A graphical representation of the presented framework is reported in Fig. \ref{fig:repr}.

\begin{remark}
While fixed final conditions are common in OCP formulations, we ignore them in this work. The described framework allows to impose such conditions by constraining $\bm{x}^-_{\na+1}$; in this work we constrain final conditions  by means of continuous-time path constraints.
\end{remark}

The continuously differentiable function $\bm{g}: \mathbb{R}^{n_x}\times \mathcal{H} \rightarrow \mathbb{R}^{n_g}$ and the function $\bm{G}: \mathbb{R}^{n_x}\times \mathbb{R}^{n_v} \rightarrow \mathbb{R}^{n_G}$ are the continuous-time and discrete-time inequality constraints functions; path constraints are respectively satisfied imposing $\bm{g}(\bm{x}_i, t) \preceq 0, \; t \in \mathcal{H}_i, \forall i \in \mathcal{I}^{\na}$, and $\bm{G}\left(\bm{x}^{-}_i, \bm{\Delta v}_i\right) \preceq 0, \; \forall i \in \mathcal{I}^{\na+1}$.\\
Finally, the continuously differentiable functions $L:\mathbb{R}^{\nx}\times \mathbb{R}_+ \rightarrow \mathbb{R}$, $\mathcal{L}:\mathbb{R}^{n_v}\rightarrow \mathbb{R}$ are the terminal and discrete running costs.

Gathering the impulses in $\bm{\Delta v} \in \mathbb{R}^{n_v\times(\na+1)}$, the subintervals bounds in $\bm{t}\in \mathbb{R}^{\na+1}$ and the state trajectories and final state in $\bm{\chi}\in \mathbb{R}^{\nx \times (\na+1)}$, the final OCP reads


\begin{problem}
\begin{equation}
    \arraycolsep=3.5pt
    \begin{array}{rl}
    \underset{\displaystyle \bm{\chi}, \bm{t}, \bm{\Delta v}}{\text{minimize}} & \displaystyle \hspace{1.5ex}L\left(\bm{x}^-_{\na + 1}, t_{\na+1}\right) + \sum_{i=1}^{\na+1}\mathcal{L}(\bm{\Delta v}_i)\\
    \text{subject to} & \left|\begin{array}{l}
    \left.\begin{array}{l}
    \dot{\bm{x}}_i = \bm{f}(\bm{x}_i, t) \\[0.5ex]
    \bm{g}(\bm{x}_i, t) \preceq 0  \\ 
    \end{array} \right|\; t \in \mathcal{H}_i, \,\forall i \in \mathcal{I}^{\na} \\[2.0ex]
    
    \;\,\bm{x}^-_{i+1} = \bm{x}^+_{i} + B \bm{\Delta v}_{i+1}  \quad \forall i \in \mathcal{I}^{\na}\\[0.5ex]
    \;\,\bm{G}(\bm{x}^-_i, \bm{\Delta v}_i) \preceq 0  \qquad \quad\;\,  \forall i \in \mathcal{I}^{\na+1}  \\[0.5ex]
    \;\,\bm{x}^-_{1} = \bm{x}_{\text{i}} + B \bm{\Delta v}_{1} \\[0.5ex]
        \;\, t_1 = \ti     \\[0.5ex] 
                                \end{array}\right. \\
                                
    \end{array}
\end{equation}
\label{prob:1}
\end{problem}

\section{TIME DILATION, CONSTRAINT AUGMENTATION, AND SCP FRAMEWORK }
\label{sec:tconstraug}

\subsection{Generalized Time Dilation}\label{subsec:dilation}

Define the continuously differentiable mapping $t : [0,\na]\rightarrow \mathbb{R}_+$ and an additional continuous control input, the \textit{dilation factor} $s : [0,\na]\rightarrow \mathbb{R}_{++}$. Defining of the \textit{dilated time} $\tau\in[0,\na]$, $t, \tau$ and $s$ are linked by the derivative map
\begin{equation}
    s(\tau) = \dfrac{\text{d}t(\tau)}{\text{d}\tau} = {t}'(\tau) 
\end{equation}
where the notation $(\cdot)'$ denotes the derivative with respect to $\tau$. $t$ can be hence treated as an additional state. As before, we divide the dilated time horizon $\bm{H}\coloneqq[0,\na]$ into $\na$ arcs $\bm{H}_i\coloneqq[\tau_i,\tau_{i+1}], \, i \in \mathcal{I}^{n_a}$. Using chain rule, dynamics are reformulated as follows.\\
\begin{equation}
\label{eq:dyn_aug}
\left[\begin{array}{c}
\bm{x}_i'\\
{t}_i'\\
\end{array}\right] = \left[\begin{array}{c}
\dot{\bm{x}}_i\\
1
\end{array}\right] \dfrac{\text{d}t_i}{\text{d}\tau} = \left[\begin{array}{c}
\dot{\bm{x}}_i\\
1
\end{array}\right] s_i, \quad  \forall \, i \in \mathcal{I}^{\na} 
\end{equation}

We fix the dilated times to the arbitrary values $\tau_i=i-1, \, i \in \{1,\dots, \na+1\}$; dilation factors render physical times free. 

\subsection{Constraint reformulation}
\label{subsec:constr}
Consider the \textit{exterior penalty functions} $q_{j,i} : \mathbb{R} \rightarrow \mathbb{R}_+,~~j=1,\dots,n_g, \, i \in \mathcal{I}^{\na}$,  obeying the constraints
\begin{equation}
\label{eq:ext_pen}
q_{j,i}(z)  \begin{cases}
                    {}= 0 &\text{if }z \le 0 \\
                    {}> 0 & \text{otherwise}
                \end{cases}
\end{equation}
and the \textit{penalty functions} $\Lambda_i: \mathbb{R}^{n_x} \times \bm{H}_i \rightarrow \mathbb{R}_+$
\begin{equation}
\label{eq:pen}
    \Lambda_i(\bm{x}_i, t_i) \coloneqq \sum_{j=1}^{n_g} q_{j,i}(g_j(\bm{x}_i, t_i))
\end{equation}
where $g_j$ denotes the scalar-valued elements of $\bm{g}$.

\begin{remark}
    The functions $q_{j,i}$ and $\Lambda_{i}$ can be tailored to each interval $\bm{H}_i$. Use of different penalty functions does not invalidate the subsequent discussion. 
\end{remark}

\begin{thm}
Path constraints are satisfied continuously over the full control horizon, i.e. for $\tau \in \bm{H}_i \;\forall \, i \in \mathcal{I}^{\na}$, if and only if there exists $y : [0,\na]\rightarrow \mathbb{R}_+$ solving the following set of initial value problems
\begin{equation}
\arraycolsep=1.5pt
\left.\begin{array}{rl}
    {y}_i'(\tau) = & s_i(\tau)\Lambda(\bm{x}_i(\tau),t_i(\tau)),\quad \tau \in \bm{H}_i\\[0.5ex]
y_i(\tau_i) = & y_i(\tau_{i+1})
\end{array}\;\right|\;\forall\,i \in \mathcal{I}^{\na}
\label{eqth:1}
\end{equation}
\end{thm}
\vspace{0.3cm}
\begin{pf}
Using Eqs. \eqref{eq:ext_pen}, \eqref{eq:pen}, the condition
$\bm{g}(\bm{x}_i(\tau), t_i(\tau)) \preceq 0, \; \tau \in \bm{H}_i \; \forall i \in \mathcal{I}^{\na}$ is equivalent to 
\begin{equation}
\label{eqpf:1}
\begin{array}{c}
\Lambda(\bm{x}_i(\tau), t_i(\tau)) = 0, \quad \tau \in \bm{H}_i, \;\forall i \in \mathcal{I}^{\na}
\end{array}
\end{equation}
Furthermore, Eq. \eqref{eqpf:1} is equivalent to the following set of \textit{isoperimetric constraints}:    
\begin{equation}
\label{eqpf:2}
\begin{array}{c}
\displaystyle
\int_{\tau_i}^{\tau_{i+1}}s_i(\tau)\Lambda(\bm{x}_i(\tau), t_i(\tau)) \text{d}\tau = 0, \qquad \forall i \in \mathcal{I}^{\na}
\end{array}
\end{equation}
Indeed, Eq. \eqref{eqpf:1} directly implies Eq. \eqref{eqpf:2}. On the other hand, the mapping $\tau \in \bm{H}_i\mapsto s_i(\tau)\Lambda(\bm{x}_i(\tau), t_i(\tau))$ is continuous since $\bm{g}$, $s$ and $\bm{x}_i, t_i$ evolving according to Eq. \eqref{eq:dyn_aug} over $\bm{H}_i$ $\forall i \in \mathcal{I}^{\na}$, are all continuous. In addition, $\Lambda$ and $s$ are respectively nonnegative and positive. Therefore Eq. \eqref{eqpf:2} necessarily requires Eq. \eqref{eqpf:1} to hold. Finally, let $y : [0,\na]\rightarrow \mathbb{R}_+$ evolve according to the dynamics in Eq. \eqref{eqth:1}, with initial state $y_1(\tau_1)=0$. Since $s_i(\tau)\Lambda(\bm{x}_i(\tau),t_i(\tau))$ is continuous, hence Riemann-integrable, $y_i(\tau), \; \tau \in \bm{H}_i\;\;\forall i \in \mathcal{I}^{\na}$ obeys 
\begin{equation}
    y_i(\tau_{i+1}) = y_i(\tau_i) + \int_{\tau_i}^{\tau_{i+1}} s_i(\tau)\Lambda(\bm{x}_i(\tau),t_i(\tau)) \text{d}\tau
    \label{eqpf:3}
\end{equation}
Therefore, by Eq. \eqref{eqpf:3}, boundary conditions in Eq. \eqref{eqth:1} are satisfied by $y$ if and only if condition in Eq.  \eqref{eqpf:2} is satisfied. \\
Satisfaction of path constraints over the full control horizon is then equivalent to existence of a mapping $y$ satisfying the set of initial value problems in Eq. \eqref{eqth:1}.
\qed
\end{pf}
\begin{remark}
    Equality path constraints can be incorporated into $\Lambda$ with appropriate additional exterior penalty functions; the described proofs remain valid under mild regularity assumptions \cite{Elango2024-dd}.
\end{remark}

\begin{remark}
As a gradient-based optimization algorithm will be used, it is of stark importance to pick continuously differentiable exterior penalty functions, e.g.
    \begin{equation}
        \arraycolsep=1.5pt
    \begin{array}{rl}
        q_j(z) &= \text{max}\{0, z\}^n \\[1ex] \Lambda(\bm{x}_i, t_i) &= \sum_{j=1}^{n_g}\text{max}\{0, g_j(\bm{x}_i, t_i)\}^n, \quad n>1
        \end{array}
    \end{equation}
\end{remark}

At this stage, it is possible to finalize reformulation of Problem \ref{prob:1}. Let us consider the augmented state $\tilde{\bm{x}}_i \in \mathbb{R}^{\nx+2}$, defined over $\bm{H}_i$ as
\begin{equation}
    \tilde{\bm{x}}_i = \left[\bm{x}_i^\text{T}, y_i, t_i\right]^\text{T}
\end{equation}
Similarly as what done before, from use of chain rule it results
\begin{equation}
\tilde{\bm{x}}'_i\ =  \left[\begin{array}{c}
\bm{f}(\bm{x}_i, t_i)\\
\Lambda(\bm{x}_i, t_i) \\
1
\end{array}\right] s_i \coloneqq \bm{F}(\tilde{\bm{x}}_i, s_i) \quad  \forall \, i \in \mathcal{I}^{\na} 
\end{equation}

Gather the dilation factors into $\bm{s} \coloneqq \left[s_1, s_2, \dots, s_{\na}\right]^{\text{T}}$, the initial conditions $\bm{x}_\text{i}, \; y_{\text{i}}\coloneqq 0$ and $t_\text{i}$ into $\tilde{\bm{x}}_\text{i} = [\bm{x}^{\text{T}}_\text{i}, y_{\text{i}}, t_\text{i}] ^{\text{T}}$. Furthermore,  $\tilde{B}\in \mathbb{R}^{(\nx+2) \times n_v}$ defines the augmented allocation matrix, $\tilde{\bm{\chi}}\in \mathbb{R}^{(\nx+2) \times (\na+1)}$ gathers the augmented state trajectories and augmented final state and $\tilde{L}:\mathbb{R}^{\nx +2}\rightarrow \mathbb{R}$ generalizes $L$ according to $\tilde{L}(\tilde{\bm{x}}) \coloneqq L(\selector{x}\tilde{\bm{x}}, \,\selector{t}\tilde{\bm{x}})$.  The original OCP gets then rearranged as

\begin{problem}
\begin{equation}
    \arraycolsep=3.5pt
    \hspace{-0.15cm}\begin{array}{rl}
    \underset{\displaystyle \tilde{\bm{\chi}}, \bm{\Delta v}, \bm{s}}{\text{minimize}} & \displaystyle \hspace{1.5ex}\tilde{L}\left(\tilde{\bm{x}}^-_{\na + 1}\right) + \sum_{i=1}^{\na+1}\mathcal{L}(\bm{\Delta v}_i)\\[3ex]
    \text{subject to} & \left|\begin{array}{l}
    \left.\begin{array}{l}
    \tilde{\bm{x}}'_i = \bm{F}(\tilde{\bm{x}}_i, s_i) \\[0.5ex]
    s_i > 0 \\[0.5ex] \selector{y} (\tilde{\bm{x}}_{i+1} - \tilde{\bm{x}}_{i}) = 0 \\[0.5ex]
    \end{array} \right|\; \forall i \in \mathcal{I}^{\na} \\[4.0ex]
    
    \;\,\tilde{\bm{x}}^-_{i+1} =\, \tilde{\bm{x}}^+_{i} + \tilde{B} \bm{\Delta v}_{i+1} \;\;\; \forall i \in \mathcal{I}^{\na}\\[0.5ex]
    \;\,\bm{G}(\selector{x}\tilde{\bm{x}}^-_{i}, \bm{\Delta v}_i) \preceq 0  \qquad \; \forall i \in \mathcal{I}^{\na+1
    }  \\[0.5ex]
    \;\,\tilde{\bm{x}}^-_{1} = \tilde{\bm{x}}_{\text{i}} + \tilde{B} \bm{\Delta v}_{1} \\[0.5ex] 
                                \end{array}\right. \\
                                
    \end{array}
\end{equation}
\label{prob:2}
\end{problem}

\subsection{Discretization, penalization and SCP}
\label{subsec:fin_SCP}
Parameterization and discretization transform the problem in a finite-dimensional one. As first step, we \textit{parameterize} the controls $s_i$: we approximate each function $s_i$ as a linear combination of $n_S$ basis functions $\Gamma_{j,i}:\bm{H}_i\rightarrow\mathbb{R}, j=1,\dots, n_S$ of coefficients $\bm{S}_i\in\mathbb{R}^{n_S}$. By choosing $\Gamma_{j,i}, \Gamma_{j,i+1}$ that differ only for a domain translation of $\tau_{i+1}-\tau_{i}$, the dependency on $i$ drops and it results
\begin{equation}
    \begin{array}{l}\nu(\bm{S}_i, \tau) \coloneqq [\Gamma_1(\tau), \dots, \Gamma_{n_S}(\tau)]\bm{S}_i \\[0.5ex] s_i(\tau) = \nu(\bm{S}_i, \tau)\end{array}  \left|\begin{array}{l} \tau \in \bm{H}_i \\[0.5ex]
    \forall i \in \mathcal{I}^{\na} \end{array} \right.
\end{equation}
Followingly, we \textit{discretize} each arc into $K$ subarcs of equal length with $K+1$ nodes; subarcs and nodes are respectively indexed by $\mathcal{I}^{K}$ and $\mathcal{I}^{K+1}$. We replace each variable $\tilde{\bm{x}}_i$ with $K+1$ augmented state variables $\tilde{\bm{x}}^k_i, k \in \mathcal{I}^{K+1}$; then we enforce dynamical feasibility via \textit{multiple-shooting}, \textit{i.e.} by integrating dynamics on time. Defining $\tau_i^k\coloneqq \tau_i + (k-1)/K$, the following constraints are added
\begin{equation}
    \arraycolsep=1.5pt
\begin{array}{rcl}
    \tilde{\bm{x}}^1_i &=& \tilde{\bm{x}}^-_i \\
    
    \tilde{\bm{x}}^{k+1}_{i} &=& \displaystyle \tilde{\bm{x}}^{k}_{i} + \int_{\tau_i^{k} }^{\tau_i^{k+1}}\bm{F}(\tilde{\bm{x}}^k_i, \nu(\bm{S}_i, \tau))\text{d}\tau\quad \forall k \in\mathcal{I}^{K} \\
    \tilde{\bm{x}}^{K+1}_i &=& \tilde{\bm{x}}^+_i
    \end{array}
\end{equation}
Introducing the mapping 
\begin{multline}
\bm{F}_{i,k}: (\tilde{\bm{x}}^k_i, \tilde{\bm{x}}^{k+1}_i, \bm{S}_i)\mapsto\\\mapsto \tilde{\bm{x}}^{k+1}_{i} - \displaystyle \tilde{\bm{x}}^{k}_{i} - \int_{\tau_i^{k-1} }^{\tau_i^{k}}\bm{F}(\tilde{\bm{x}}^k_i, \nu(\bm{S}_i, \tau))\text{d}\tau\end{multline} dynamical feasibility can be imposed as $\bm{F}_{i,k} = 0, \forall i \in \mathcal{I}^{\na}, \forall k \in \mathcal{I}^K$. Discretization of the boundary conditions on $y$ leads instead to $\selector{y} (\tilde{\bm{x}}^{k+1}_{i} - \tilde{\bm{x}}^k_{i}) = 0, \forall i \in \mathcal{I}^{\na}, \forall k \in \mathcal{I}^K$.

\begin{remark}
 At convergence, constraints violate the linear independence constraint qualification \cite{Elango2024-dd}, a typical requirement of numerical optimization algorithms \cite{Nocedal2006-lo}. We avoid this pathological scenario by relaxing the boundary conditions on $y$ as
    $$
        \selector{y} (\tilde{\bm{x}}^{k+1}_{i} - \tilde{\bm{x}}^k_{i}) < \varepsilon, \quad\forall i \in \mathcal{I}^{\na}, \forall k \in \mathcal{I}^K
    $$
    given a small scalar value $\varepsilon$. As shown later in the results, this relaxation does not negatively affect the quality of the constraint satisfaction.
\end{remark}

The remaining constraints retain the same form as in Problem \ref{prob:2}.

\begin{remark}
    Whilst sensitivity to coarse initial guesses is mitigated by use of multiple-shooting, such last technique adds variables to the solution process; thus increasing computational times. For this reason, \textit{single-shooting} is sufficient and beneficial when sensitivity issues are not encountered.
\end{remark}

We then proceed by \textit{exactly penalizing} the nonconvex constraints, \textit{i.e.} we incorporate them in the objective function using an $\ell_1$ penalization. Let us gather the augmented state trajectories $\tilde{x}^k_{i}, i \in \mathcal{I}^{\na}, k \in \mathcal{I}^{K+1}$, the parameterized controls $\bm{S}_i, i \in \mathcal{I}^{\na}$ and the impulsive controls $\bm{\Delta v}_i, i \in \mathcal{I}^{\na + 1}$ in the single vector $\bm{z}$; moreover, let the set $\mathcal{Z}$ contain the augmented states and dilation factors feasible with respect to the convex constraints of the discretized problem, and let $\tilde{\mathcal{I}}_\mathcal{Z}$ be its indicator function. We introduce the objective function $\Theta_\gamma:\mathbb{R}^{n_z}\rightarrow\mathbb{R}$ and define it as
\begin{multline}
    \Theta_\gamma(\bm{z}) \coloneqq \tilde{L}\left(\tilde{\bm{x}}^-_{\na + 1}\right) + \sum_{i=1}^{\na+1}\mathcal{L}(\bm{\Delta v}_i) + \tilde{\mathcal{I}}_\mathcal{Z}(\bm{z}) + \\ + \gamma \sum_{i=1}^{\na + 1} \text{max}\{0, \bm{G}(\selector{x}\tilde{\bm{x}}^-_{i}, \bm{\Delta v}_i)\} + \\ + \gamma \sum_{i=1}^{\na + 1}\sum_{k=1}^{K+1}\|\bm{F}_{i,k} (\tilde{\bm{x}}^k_i, \tilde{\bm{x}}^{k+1}_i, \bm{S}_i)\|_1
\end{multline}
where $\gamma \in \mathbb{R}_+$ is a large finite value. We have exactly penalized the violation of nonconvex constraints by means of nonsmooth penalty functions, as the $\ell_1$-norm. Therefore, for a finite high value of $\gamma$, the solution of the following unconstrained problem
\begin{problem}
    \begin{equation}
    \begin{array}{rl}
    \underset{\displaystyle \bm{z}}{\text{minimize}} & \Theta_\gamma(\bm{z}) \\
    \end{array}
    \end{equation}
    \label{prob:unc}
\end{problem}
 satisfies the Karush--Kuhn--Tucker (KKT) conditions of the original constrained problem and is a strict local minimizer of it \cite{Nocedal2006-lo}. 
No approximation has been introduced up to here, except for the relaxation of the boundary conditions on $y$. This means that, solving Problem \ref{prob:unc}, we can retrieve the exact solution of Problem \ref{prob:1}.

Problem \ref{prob:unc}, however, is still nonlinear. We therefore use \textit{prox-linear} method, an SCP algorithm that minimizes convex-composite functions with guaranteed convergence. More specifically, prox-linear treats objective functions of type $\Theta_\gamma(\bm{z}) = J(\bm{z}) + H(\bm{c}(\bm{z}))$, where $J$ is a proper closed convex function, $H$ is an $\alpha$-Lipschitz continuous convex function and $\bm{c}$ is a potentially nonconvex and continuously differentiable function with $\beta$-Lipschitz continuous gradient. With respect to our formulation, $J$ corresponds to the indicator function $\tilde{\mathcal{I}}_\mathcal{Z}$; $\bm{c}$ gathers a) the objective function, b) the discrete path constraints and c) the shooting constraints. Finally, $H$ components are a) the identity function, for the objective function, b) the function $\text{max}\{0, (\cdot)\}$. for the discrete path constraints and c) the $\ell_1$-norm, for each shooting constraint.

Given the current iterate $\bm{z}_{j}$, prox-linear minimizes iteratively the following convex approximation of $\Theta_\gamma$
\begin{multline}
    \Theta^\rho_\gamma(\bm{z}, \bm{z}_{j}) = J(\bm{z}) + H\left(\bm{c}(\bm{z}_{j})+\right. \\ +\left.\nabla \bm{c}(\bm{z}_{j})(\bm{z} - \bm{z}_j) \right) + \dfrac{1}{2\rho}\left\|\bm{z} - \bm{z}_{j}\right\|_2
\end{multline}
where $\rho \in \mathbb{R}_+$ is a tuning parameter that influences the proximal term weight. Defining the \textit{prox-gradient mapping} as $\mathcal{G}_{\rho}:\bm{z}\mapsto\mathbb{R}$ as \begin{equation}\mathcal{G}_{\rho}(\bm{z}) \coloneqq \dfrac{1}{\rho}\left(\bm{z} - \underset{\tilde{\bm{z}}}{\operatorname{argmin}\;}\Theta_{\gamma}^{\rho}(\tilde{\bm{z}}, \bm{z}_j) \right)\end{equation} 
the following condition is verified at each iteration of the algorithm \cite{Drusvyatskiy2018-di}
\begin{equation}
    \Theta_\gamma(\bm{z}_{j+1}) \leq \Theta_\gamma(\bm{z}_{j}) +\left(\alpha\beta-\dfrac{1}{\rho}\right)\left\|\rho\,\mathcal{G}_{\rho}(\bm{z}_j)\right\|^2,
    \label{eq:mon_decr}
\end{equation}
Therefore, the condition $\rho \leq 1/(\alpha\beta)$ is sufficient to ensure that the objective function decreases monotonically. At last, consider the initial guess $\bm{z}_1$, a lower bound $\Theta^\star_\gamma$ for $\Theta_\gamma$ and a stopping tolerance $\epsilon$. If $\rho \leq 1/(\alpha\beta)$, the number of iterations required to bring $\left\|\mathcal{G}_\rho \right\|^2$ within some tolerance $\epsilon$ is bounded. More specifically, 
\begin{equation}
j \geq \dfrac{2 \alpha \beta (\Theta_\gamma(\bm{z}_1) -\Theta^\star_\gamma)}{\epsilon} \;\Rightarrow \; \left\|\mathcal{G}_\rho(\bm{z}_j) \right\|^2 < \epsilon 
\end{equation}
In practical applications, a maximum number of iterations $j_\submax$ is added to prox-linear for contingency scenarios. Algorithm \ref{alg:proxlinear} synthesizes steps of prox-linear.

A local minimum $\bm{z}^\star$ computed with prox-linear is close to a point where first-order optimality of $\Theta_\gamma(\bm{z})$ is small \cite{Drusvyatskiy2018-di}. This point in turn, for high finite $\gamma$, is close to a strict local minimizer of Problem \ref{prob:1}.

\begin{remark}
    $\gamma$ and $\rho$ are tuning parameters and, as such, choosing them properly is crucial to ensure algorithm convergence. $\gamma$ is heuristically selected to grant satisfactory results; backtracking schemes are employed to update $\rho$ \cite{Drusvyatskiy2018-di} and grant successful convergence to a local minimum. 
\end{remark}

\begin{algorithm}[!htpb]
\caption{Prox-linear Method}
\label{prox-linear}
\begin{algorithmic}[0]
\Require $j_{\submax}$, $\epsilon$, $\rho$
\State \hspace{-0.33cm}\textbf{Initialize:} $\bm{z}_1$
\State $j\gets 1$
\While{$j \le j_{\max}$ \textbf{and} $\left\|\mathcal{G}_\rho(\bm{z}_j)\right\| > \epsilon$}
\State $\bm{z}_{j+1} \leftarrow \underset{z}{\operatorname{argmin}}~\Theta_\gamma^\rho(\bm{z},\bm{z}_j)\phantom{X^{X^X}}$
\State $j \leftarrow j+1$
\EndWhile
\Ensure $\bm{z}_j$
\end{algorithmic}
\label{alg:proxlinear}
\end{algorithm}

\section{NUMERICAL EXAMPLE}
\label{sec:num}
\subsection{Problem definition}
\label{subsec:proboutline}
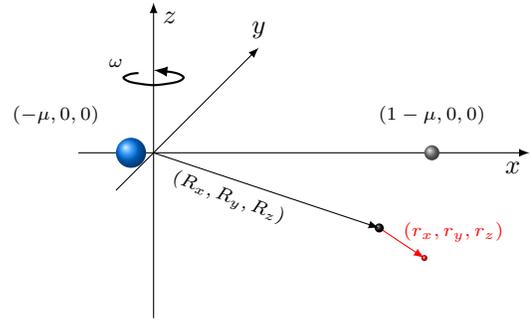
\begin{figure}[h!]
\centering
\begin{tikzpicture}[x={(1cm,0cm)}, y={(0cm,1cm)}, z={(0.5cm,0.5cm)}, scale=2]
\draw[-latex] (-0.5,0,0) -- (2.5,0,0) node[anchor=north east]{$x$};
\draw[-latex] (0,-1.1,0) -- (0,1.0,0) node[anchor=north west]{$z$};
\draw[-latex] (0,0,-0.5) -- (0,0,1.4) node[anchor=south]{$y$};

\coordinate (mu) at (-0.15,0,0);
\coordinate (mu_label) at (-0.65,0.25,0);
\coordinate (1-mu) at (1.85,0,0);
\coordinate (1-mu_label) at (1.85,0.25,0);
\coordinate (r) at (1,-1,1);
\coordinate (r_c) at (1.3,-1.2,1);
\coordinate (r_label) at (1,-1,1);
\draw[-latex, red] (r) -- (r_c); 

\shade[ball color=RoyalBlue] (mu) circle (0.1);
\shade[ball color=gray] (1-mu) circle (0.05);
\shade[ball color=black] (r) circle (0.03);
\shade[ball color=red] (r_c) circle (0.02);

\node at (mu_label) {\scriptsize $(-\mu,0,0)$};
\node at (1-mu_label) {\scriptsize$(1 - \mu,0,0)$};
\node[rotate=-20] at (0.5, -0.3, 0) {\scriptsize $(R_x,R_y,R_z)$};
\draw[-latex] (0,0,0) -- (r); 
\node[above right, color=red] at (1.1, -1.15, 1) {\scriptsize $(r_x,r_y,r_z)$};
    \draw[latex-, thick] plot [smooth] coordinates {(0, 0.55, 0) (0.15, 0.53, 0)(0.2, 0.5, 0) (0.15, 0.47, 0)(0, 0.45, 0) (-0.15, 0.47, 0)(-0.2, 0.5, 0)  (-0.15, 0.525, 0) (-0.1, 0.53, 0)};
\node at (-0.25, 0.6, 0) {\scriptsize $\omega$};
\end{tikzpicture}
\caption{Earth, Moon, target and chaser spacecrafts (in black and red) in the normalized Earth-Moon synodic reference frame.}
\label{fig:CR3BP}
\end{figure}

We assume a target on a 9:2 resonant southern NRHO emanating from $L_2$ in the Circular Restricted 3 Body (CR3B) model. Motion of each spacecraft is described in the Earth-Moon synodic reference frame; the absolute dynamics RHS, normalized with respect to mean Earth-Moon distance and angular speed $\omega$, reads
\begin{equation}
    \bm{f}^a = \left[\begin{array}{c}
        \bm{V} \\
        R_x + 2V_y -(1-\mu)\dfrac{R_{1,x}}{R_1^3} -\mu\dfrac{ R_{2,x}}{R_2^3} \\[2ex] 
        R_y - 2V_x -(1-\mu)\dfrac{R_{1,y}}{R_1^3} -\mu\dfrac{ R_{2,y}}{R_2^3} \\[2ex]
        -(1-\mu)\dfrac{R_{1,z}}{R_1^3} -\mu\dfrac{ R_{2,z}}{R_2^3}
    \end{array}\right]
\end{equation}
where $\bm{R}_1$ and $\bm{R}_2$ are the vectorial relative distances from Earth and Moon, and $R_1$ and $R_2$ their norms.
Fig. \ref{fig:CR3BP} shows the target and the chaser in the synodic Earth-Moon reference frame. 

 Initial state of the target is fixed at perilune: the function $\bm{X}_\text{T}(t)$ that describes the target state likewise parameterizes in time the NRHO on which the target resides. The chaser can fire $\na$ impulses, each of maximum magnitude $\Delta v_\submax$. In addition, the target must remain inside a sphere of radius $r_\submax$ and outside a sphere of radius $r_\submin$. Both spheres are centered at the target.
The objective is maximizing the residence time of the chaser in the admissible region, given the available control actuation capability. Problem is hence formulated as
\begin{problem}
\begin{equation}
    \arraycolsep=3.5pt
    \hspace{-0.2cm}\begin{array}{rl}
    \underset{\displaystyle \bm{\chi}, \bm{t}, \bm{\Delta v}}{\text{minimize}} & \displaystyle \hspace{1.5ex}-t_{\na+1}\\
    \text{subject to} & \left|\begin{array}{l}
    \left.\begin{array}{l}
    \dot{\bm{x}}_i = \bm{f}(\bm{x}_i, t) \\[0.5ex]
    \|\selector{r}\bm{x}_i\|_2 - r_\submax \leq 0  \\[0.5ex] 
       r_\submin - \|\selector{r}\bm{x}_i\|_2  \leq 0  \\ 
    \end{array} \right| \begin{array}{l} t \in \mathcal{H}_i \\ \forall i \in \mathcal{I}^{\na} \end{array} \\[4.0ex]
    \;\,\bm{x}^-_{i+1} = \bm{x}^+_{i} + B \bm{\Delta v}_{i+1}  \quad \forall i \in \mathcal{I}^{\na}\\[0.5ex]
    \;\,\|\bm{\Delta v}_i\|_2\leq \Delta v_\submax  \qquad \quad\;\,  \forall i \in \mathcal{I}^{\na+1}  \\[0.5ex]
    \;\,\bm{x}^-_{1} = \bm{x}_{\text{i}} + B \bm{\Delta v}_{1} \\[0.5ex]
        \;\, t_1 = 0     \\[0.5ex] 
                                \end{array}\right. \\
                                
    \end{array}
\end{equation}
\label{prob:4}
\end{problem}

\subsection{Numerical results}
\label{subsec:results}

A two and a three-impulse strategy are tested. The devised two-impulse strategy consists of an initial free-drift arc, followed by two impulses; the three-impulse strategy allows an initial impulse as well. In both cases, an initial position deviation of 400 m is accounted for; for the two-impulse strategy, deviation is along the $y$ axis, for the two impulse, it is along the $x$ axis. Common problem parameters are reported in Tab. \ref{tab:data}. In both cases, a First-Order Hold parameterization for the dilation factor is used.

\begin{figure}
    \centering
    \begin{subfigure}[]{\linewidth}
        \centering
    \includegraphics[width=0.85\linewidth]{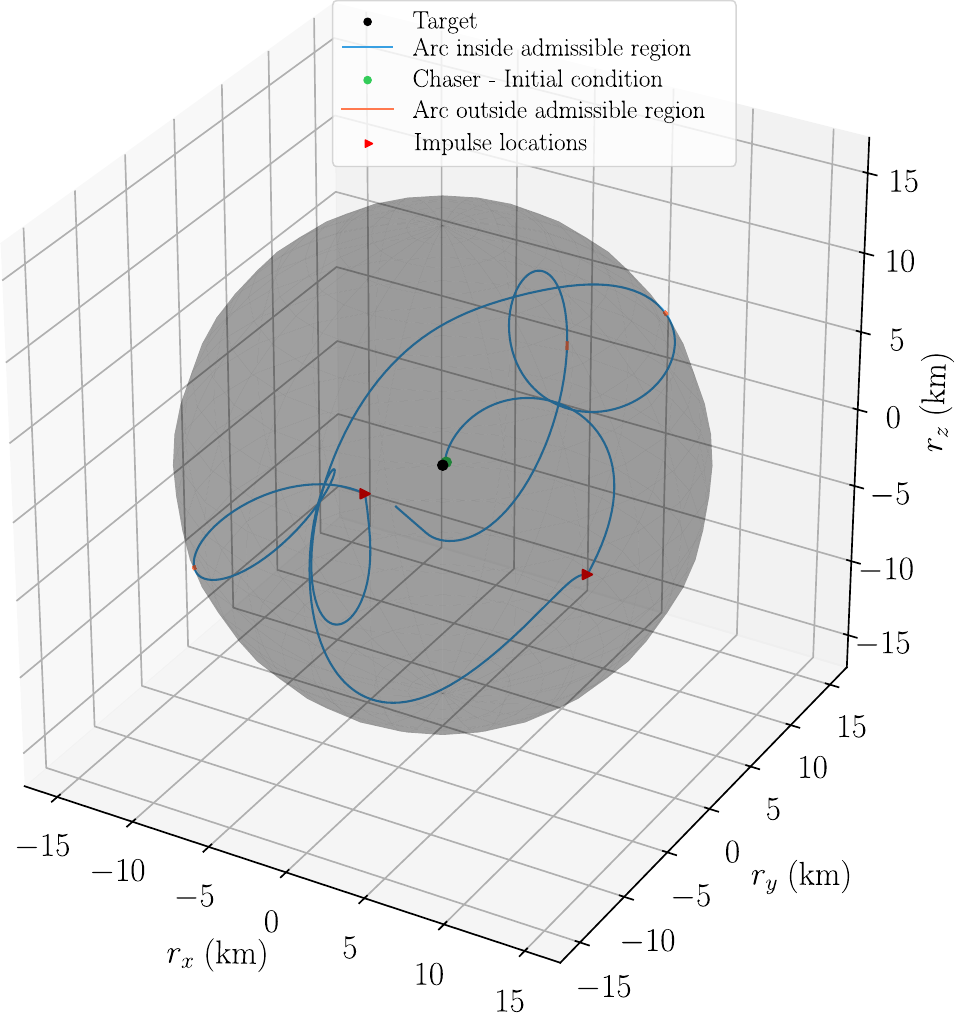}
        \caption{Coasting and 2-impulse strategy}
        \label{fig:2imp}
    \end{subfigure}
    \vspace{0.5cm}
    
    \begin{subfigure}[]{\linewidth}
        \centering
    \includegraphics[width=0.85\linewidth]{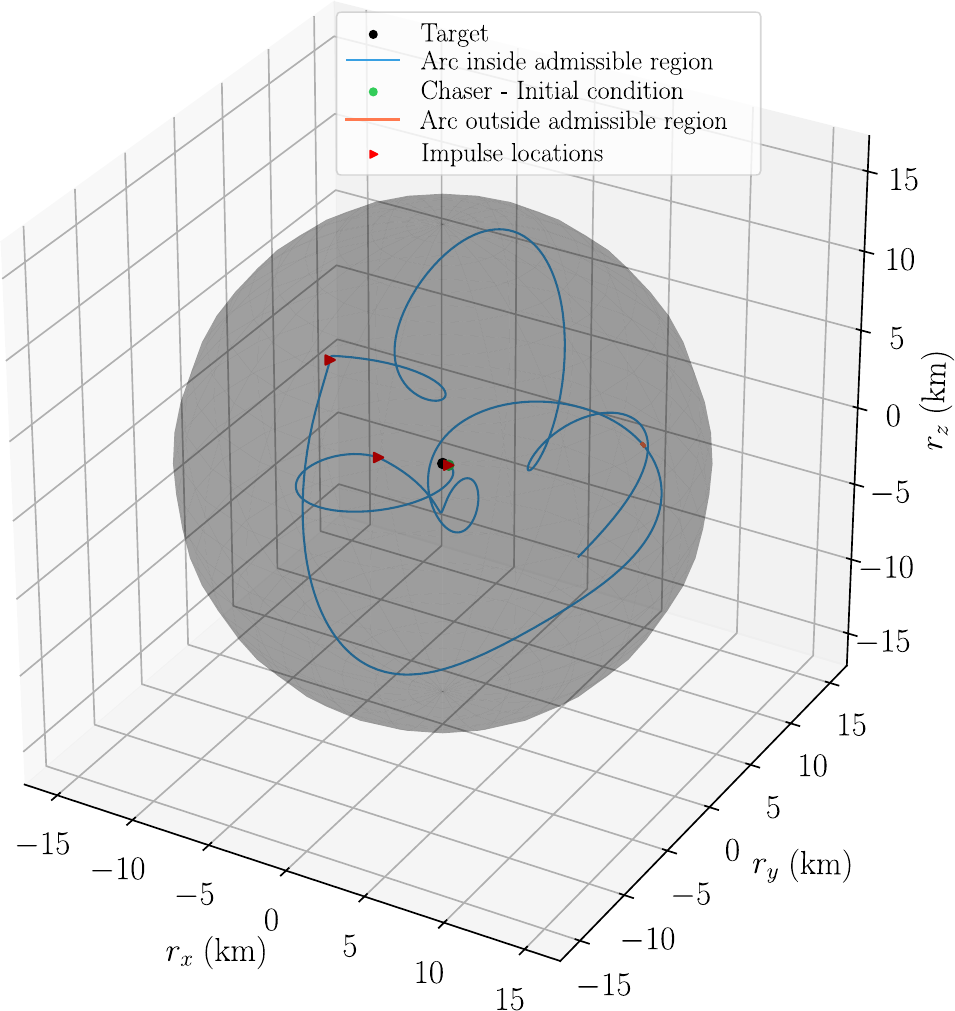} 
    \caption{3-impulse strategy}
    \label{fig:3imp}
    \end{subfigure}
    \caption{Relative trajectories in the synodic reference frame}
    \label{fig:rel_traj}
\end{figure}

\begin{table}[ht]
    \centering
    \caption{\label{tab:data} Data for the two scenarios}
    \begin{threeparttable}
    \begin{tabular}{ccc}
    \toprule
    Physical quantity &  Value & Unit \\
    \midrule
    $R_{T}(0)^\dagger$    & $[ 0.987360158,  0,   0.008773055]$ & (-) \\[0.5ex]
    $V_{T}(0)$    & $[ 0,   1.634461555,   0]$& (-)\\[0.5ex]
    $r_\submax$ & $15$ & km \\[0.5ex]
    $r_\submin$ & $0.3$ & km \\[0.5ex]
    $\Delta v_\submax $ & $2.5 \cdot10^{-4}$ & km/s \\
    [0.5ex]
    \bottomrule    
    \end{tabular}
    \begin{tablenotes}
    \item $^\dagger$ For conversion purposes, Earth-Moon mean distance is approximated to 384400 km
    \end{tablenotes}
    \end{threeparttable}
\end{table}
By letting the system drift from its initial conditions without control for one full period (6 days and a half), distance-to-target peaks at approximately 200 km for the 3-impulse case, and at 50 km for the two-impulse case; this has been verified during simulations, and is not reported for space limitations. The need for control is then evident. The final trajectories are represented in Fig. \ref{fig:rel_traj} for both strategies; small arcs of KIZ constraint violation appear in both cases; except for these, the trajectories lie completely within the admissible zone. 
\begin{figure*}[ht]
      \centering
    \begin{subfigure}[]{\linewidth}
        \centering
    \includegraphics[width=\linewidth]{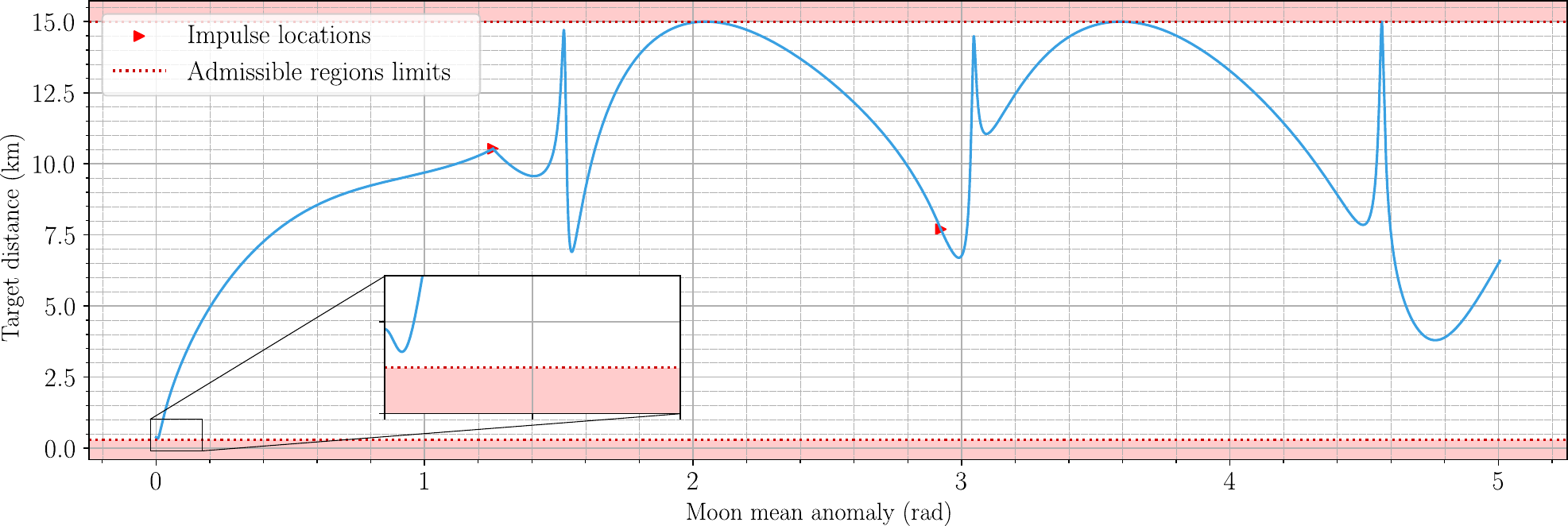}
        \caption{Coasting and 2-impulse strategy}
        \label{subfig:4a}
    \end{subfigure}
    \vspace{0.5cm}
    
    \begin{subfigure}[]{\linewidth}
        \centering
    \includegraphics[width=\linewidth]{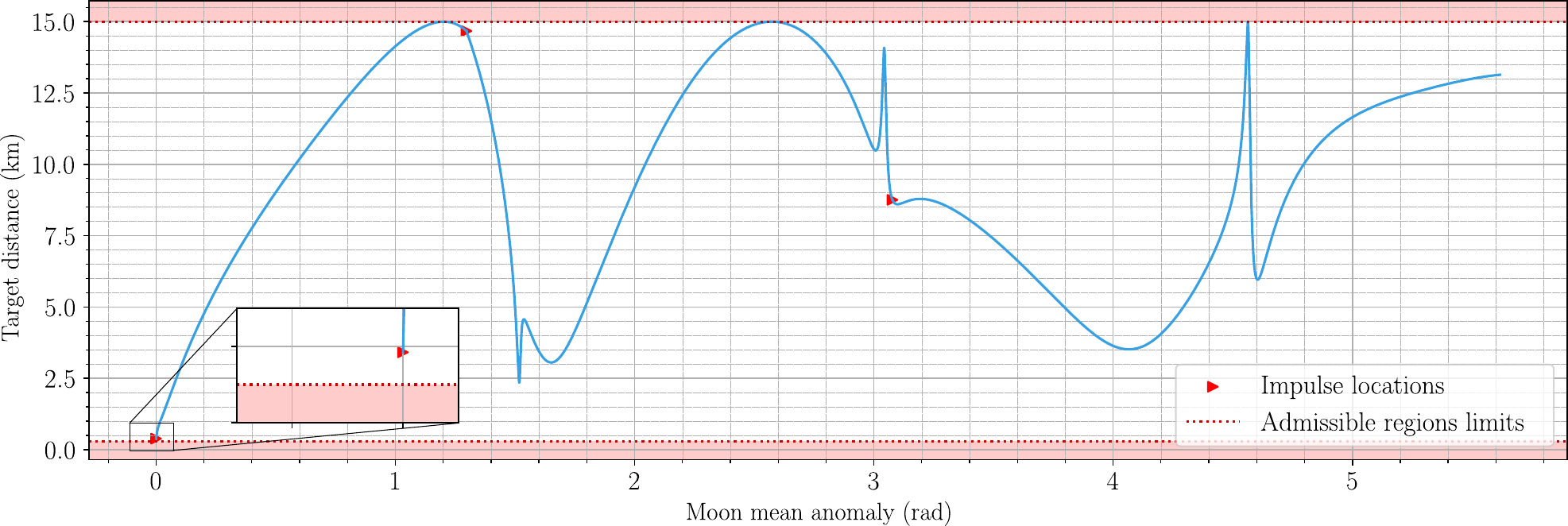} 
    \caption{3-impulse strategy}
    \end{subfigure}
    \caption{Relative-to-target distance with isoperimetric constraint reformulation}
    \label{fig:rel_dist}
\end{figure*}

Table \ref{tab:results} formalizes the results for the two case studies. 
\begin{table}[ht]
    \centering
    \caption{\label{tab:results} Merit parameters for the two scenarios}
    \begin{threeparttable}
    \begin{tabular}{ccc}
    \toprule
    \multirow{ 2}{*}{Parameter} & \multicolumn{2}{c}{Value} \\
    {} & \hspace{0.7cm}2-impulses\hspace{0.7cm}{} & \hspace{0.7cm}3-impulses\hspace{0.7cm}{} \\
    \midrule
    $t_{\na + 1}$    & 22.3 d & 25.0 d \\[0.5ex]
    $\Delta v_1$    & (-) & 19.5 cm/s \\[0.5ex]
    $\Delta v_2$    &  7.8 cm/s & 5.2 cm/s\\[0.5ex]
    $\Delta v_3$ & 7.4 cm/s & 4.9 cm/s \\[0.5ex]
    CPU time$^\dagger$/it. & 6.34 ms & 5.46 ms \\[0.5ex]
    $\text{its.}$ & 71 & 55 \\
    [0.5ex]
    CPU time & 454 ms & 301 ms \\
    [0.5ex]
    \bottomrule    
    \end{tabular}
    \begin{tablenotes}
    \item $^\dagger$ Tests have been run on Julia 1.10.3, on a MacBook 2023 with Apple M3 Pro, 18GB of unified memory. Mosek has been used as convex solver.
    \end{tablenotes}
    \end{threeparttable}
\end{table}
In both cases the proposed predictive framework grants at least 22 days of bounded loitering. A single segment for each arc is used in the discretization step: CPU times per iteration are extremely contained. The discretization time associated with integration of a single subproblem penalizes the full algorithm; such time is approximately two orders of magnitude higher than the CPU time required to solve the subproblem itself. On the other hand, this limitation is common to shooting techniques and is not due to the proposed state augmentation. Furthermore, dedicated integrators, which are not the focus of this work, can alleviate this problem. In addition, the proposed predictive control strategy ensures that a single node is sufficient to capture the full amount of violations of the path constraints, no matter the duration of the coasting arc; as shown in Fig. \ref{fig:rel_dist}, indeed, the distance relative to target remains within the acceptable region for both case studies. 
Notably, GTD influence is evident by comparing the length of the first and of the last arcs in Figs. \ref{fig:2imp}, \ref{fig:3imp}. The guess trajectory considers nodes at equally spaced times; at convergence, loitering time after the last impulse is nearly the 200\% of the loitering time of the coasting arc in Fig. \ref{fig:2imp} and of the loitering time after the first impulse in Fig. \ref{fig:3imp}. This confirms the performance increment provided by GTD. The partial constraint violation highlighted in Fig. \ref{fig:rel_traj} falls within the bounds related with the relaxation in Sec. \ref{subsec:fin_SCP}; constraint violation can be quantified and contained. More importantly, no violation of the KOZ constraint is verified at all, as highlighted by the zoom-ins in Fig. \ref{fig:rel_dist}. The proposed technique exploits shooting with a single node for each coast arc whilst ensuring constraint satisfaction over the continuous-time horizon.

\subsection{Comparison with node-only constraints}
\begin{figure*}[ht]
    \centering
    \includegraphics[width=\linewidth]{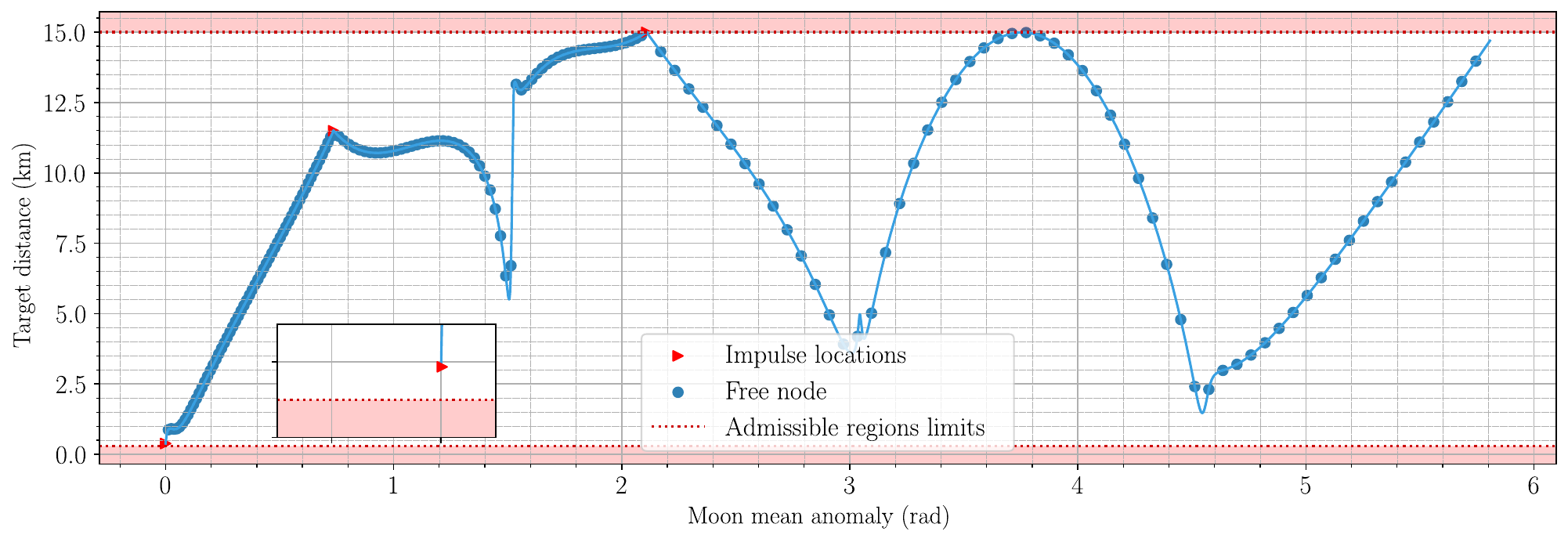}
    \caption{Relative-to-target distance for node-only constraints - 3-impulse strategy}
    \label{fig:3imp_mr}
\end{figure*}

 Mesh-refinement techniques impose constraints only at nodes and address inter-sample constraint violation indirectly, adding more nodes and thickening the discretization grid of the time horizon.
 Although intuitive, mesh-refinement techniques may require a large number of nodes to ensure constraint satisfaction over the full horizon. The three-impulse strategy is tested by imposing path constraints only at nodes, without isoperimetric reformulation. Nodes are uniformly distributed on each arc $\bm{H}_i$; the total number of nodes is the minimum number that guarantees satisfaction of the path constraints for the entire time horizon at convergence of the algorithm. Results are reported in Fig. \ref{fig:3imp_mr}. 
In the tested scenario, the total required number of nodes per arc amounts to 60, for a total number of 180 nodes. The corresponding computational performance is reported in Tab. \ref{tab:mr}.
 \begin{table}[ht]
    \centering
    \caption{\label{tab:mr} Merit parameters for node-only constraints}
    \begin{threeparttable}
    \begin{tabular}{cc}
    \toprule
    \multirow{ 2}{*}{Parameter} & Value \\
    {} & \hspace{1.4cm}3-impulses\hspace{1.4cm}{} \\
    \midrule
    $t_{\na + 1}$    & 25.9 d \\[0.5ex]
    CPU time$^\dagger$/it. & $>$ 400 ms \\[0.5ex]
    $\text{its.}$ & 17 \\
    [0.5ex]
    CPU time & $>$ 6800 ms \\
    [0.5ex]
    \bottomrule    
    \end{tabular}
    \begin{tablenotes}
    \item $^\dagger$ Tests have been run on Julia 1.10.3, on a MacBook 2023 with Apple M3 Pro, 18GB of unified memory. Mosek has been used as convex solver.
    \end{tablenotes}
    \end{threeparttable}
\end{table}
As evident by comparing Tables \ref{tab:results}, \ref{tab:mr}, the computational time required by the isoperimetric reformulation is lower than the 5\% of the time required by the traditional approach based on mesh refinement. This result is obtained for equal performance of the two approaches with respect to constraint satisfaction. Furthermore, the loitering time obtained with the standard approach is slightly higher than that obtained with the isoperimetric reformulation. This happens since the traditional approach uses more variables to solve each convex subproblem with respect to the approach with the isoperimetric reformulation: in fact, the former uses 180 state vector variables, the latter uses only 3. The gains in computational time are however evident, and prove the performance of the proposed approach for safety-critical applications. 

\label{subsec:meshrefinement}

\section{CONCLUSIONS}
\label{sec:concl}

In this work, we have proposed and tested a strategy to compute relative constrained trajectories with impulsive controls. We have optimized location and timing of impulses; with few nodes, our approach has successfully satisfied path constraints throughout the entire computed time horizon. The framework has been tailored for a specific application in the double Earth-Moon gravitational regime. A Sequential Convex Programming algorithm has been used, showing fast solution computation and adaptability to different scenarios.


\printbibliography

\end{document}